\documentclass{elsart}
\begin{document}

\runauthor{}

\begin{frontmatter}
\title{Giant magnetoresistance in quantum magnetic contacts}

\author[KSU]{L.R. Tagirov\thanksref{Someone}},
\author[KSU,KFTI]{B.P. Vodopyanov},
\author[KSU]{B.M. Garipov}

\address[KSU]{Kazan State University, Kazan 420008,
Russian Federation}
\address[KFTI]{Kazan Physico-Technical Institute of RAS, Kazan 420029,
Russian Federation}
\thanks[Someone]{Corresponding author, E-mail address: Lenar.Tagirov@ksu.ru \\
Postal address: Theoretical Physics Department, Kazan State University, 
420008 Kazan, Russia ; FAX: 007-8432-387201}

\begin{abstract}
We present calculations of quantized conductance and magnetoresistance 
in nanosize point contacts between two ferromagnetic metals. 
When conductance is open for only one conduction electrons spin-projection,
the magnitude of magnetoresistance is limited by the rate of
conduction electron spin-reversal processes. For the case when both spin-channels 
contribute to the conductance we analyze the influence of the point contact 
cross-section  asymmetry on the giant megnetoresistance. Recent experiments 
on magnetoresistance of magnetic point contacts are discussed in the 
framework of the developed theory.

\end{abstract}
\end{frontmatter}

\newpage

\section{Introduction}

Since experiments with two-dimensional electron gas in a semiconductor \cite
{Wees,Wharam} it is demonstrated that electric conduction is quantized, and
elementary conductance quantum is equal to $2e^2/h$. When measured on tiny
contacts of nonmagnetic semiconductors and metals the conductance
quantization is limited to low temperatures by thermal fluctuations, and the
factor 2 is attributed to the two-fold spin degeneracy of conduction
electron states. Recently, sharp conductance quantization steps have been
observed in nanosize point contacts of ferromagnetic metals at room
temperature \cite{Kramer,Oshima,Ott,Ono}. It was possible, because phonon
and magnon assisted relaxation processes are quenched due to a large, $\sim
1eV$, exchange splitting of the conduction band. In addition, Oshima and
Miyano \cite{Oshima} found an indication of the odd-valued number $N$ of
open conductance channels ($\sigma =N(e^2/h)$) in nickel point contacts from
room temperature up to 770K. Ono \textit{et al.} \cite{Ono} presented an
evidence of switching from $2e^2/h$ conductance quantum to $e^2/h$ quantum
at room temperature in the nickel nanocontacts of another morphology.
Obviously, the change of conductance quantum from $2e^2/h$ to $e^2/h$ is a
result of lifting-off the spin degeneracy of the conduction band. Recent
calculations \cite{Imamura,Zvezdin} confirmed the $e^2/h$ conductance
quantization in ferromagnetic metals, which is due to non-synchronous
opening of ''up'' and ''down'' spin-channels in the point contact conduction.

New pulse to studies of electric transport in ferromagnets has been given by
observation of giant magnetoresistance (GMR) in nanosize magnetic contacts
by Garc\'ia \textit{et al.} \cite{Garcia1,Garcia2,Garcia3}.
Magnetoresistance magnitudes of 280\% for Ni-Ni \cite{Garcia1} and 200\% for
Co-Co \cite{Garcia2} nanosize contacts were obtained at room temperature.
Somewhat smaller ($\sim 30\%$), but also very large for a single interface,
magnetoresistance was observed in Fe-Fe point contacts \cite{Garcia3}. In
these experiments there is a huge spread in the measured values of
magnetoresistance, drawn as a function of conductance at ferromagnetic
alignment of magnetizations in contacting ferromagnetic domains
(F-conductance). The spread of MR points for Ni-Ni and Co-Co contacts is
extremely large at F-conductance lying in the range of 2-8 elementary
conductances $e^2/h$. The above mentioned observations of conductance
quantization steps in point junctions of ferromagnetic metals at room
temperature give anticipation that conduction quantization is responsible
for the giant magnitude and the giant fluctuations of magnetoresistance in
tiny magnetic contacts.

In this article we develop a simple model of conductance and
magnetoresistance for nanosize magnetic contacts in the regime of
conductance quantization (quantum magnetic contacts), proposed in the
previous work \cite{Tagirov2}. In \cite{Tagirov2} we argued, that if only
one conduction electron spin-channel is open at F-conductance, then the
magnitude of GMR is limited only by spin-flip processes of conduction
electrons when passing through the point contact \cite{Referee}. Then the
magnetic nanocontact serves as quantum spin-valve. When both spin-channels
of F-conduction are open we established, that GMR is a multivalued function
of conductance at ferromagnetic alignment of magnetizations (at least at low
temperatures and absence of disorder). This means that if the conductance is
quantized, different samples, having the same F-conductance, reveal
different magnetoresistance. Distribution of magnetoresistance values is not
normal or flat in the statistical sense. Rather, at fixed F-conductance
values, smaller magnitudes are much more probable than the maximal ones. The
width of distribution is extremely large for the first few open
F-conductance channels. Thus, we concluded that the giant raw-data
fluctuations observed in the experiments by Garc\'ia \textit{et al.} \cite
{Garcia1,Garcia2,Garcia3} might be the consequence of conduction
quantization. In the present study we focus out attention on the influence
of the point contact cross-section asymmetry on GMR. We find that number of
open conduction channels, at which conductance shows up for the
antiferromagnetically aligned magnetizations, depends not only on the
conduction band spin-polarization \cite{Tagirov2}, but also on the aspect
ratio of the contact cross-section. We discuss the above mentioned as well
as very recent experiments on GMR in magnetic nanocontacts.

\section{Calculation of conductance and magnetoresistance}

We consider a model of two ferromagnetic, single-domain half-spaces
contacting via a narrow and short neck with typical length from one to
several nanometers. For the F-alignment of domains the magnetization is
homogeneous along the constriction, therefore current carriers move in a
constant potential created by the magnetization. At antiferromagnetically
(AF) aligned domains a domain wall (DW) is created inside the neck. Then,
the carriers move in a potential landscape created by the domain wall.
According to the general quantum-mechanical prescription, any inhomogeneity
in the potential energy landscape results in a reflection of quasiparticle
wave function, which evokes an additional electric resistance \cite{Falicov}. 
For the free DW between unconstrained domains this domain-wall resistance
is very small because the profile of DW is smooth, and free domain wall
width $\delta _0$ is large, typically in the range 15-150 nm for the strong
elemental ferromagnets like Co, Fe and Ni \cite{Hubert}. However, if DW is
created in the constriction, then the wall width $\delta $ is approximately
equal to the length of the neck $l$, which is at least an order of magnitude
shorter than $\delta _0$ \cite{Bruno}. The sharpening of DW leads to huge
enhancement of quasiparticle reflection from DW \cite{Tagirov1}, as well as
some increase of impurity scattering \cite{Levy}. When the external DC
magnetic field aligns the domains magnetizations parallel (F-alignment), it
eliminates DW and domain-wall reflection, which results in essential
decrease of resistance, \textit{i.e.} leads to GMR \cite{Tagirov1}.

Now we consider the regime of quantized conductance through the nanosize
neck. The cross-section size of the neck is assumed very small, typically
about 1 nm, so that the transverse motion of electron in the neck is
quantized. In our previous work \cite{Tagirov2} we considered the neck of
cylindrical cross-section, in this paper we give solution for the neck of
rectangular shape. The length of the neck, $l$, is considered shorter than
the electron mean free path, that is why the electron transport through the
neck is ballistic. Actually, the neck is a conducting bridge which plays the
role of a quantum filter. It selects from the continuous domain of
quasiparticle incidence angles only those, which meet the allowed (and
quantized) transverse momentum in the channel, and satisfy the energy and
momentum conservation laws. For the particular calculations of conductance
we may use the ballistic-limit versions of the formulas (14), (18) and (19)
of our work \cite{Tagirov1}: 
\begin{equation}
\sigma ^F=\sigma _{\uparrow \uparrow }+\sigma _{\downarrow \downarrow }=
\frac{e^2}h\widetilde{\sum_{m,n}}\left\{ D_{\uparrow \uparrow
}(x_{mn})+D_{\downarrow \downarrow }(x_{mn})\right\} ,  \label{eq1}
\end{equation}
\begin{equation}
\sigma ^{AF}=\frac{2e^2}h\widetilde{\sum_{m,n}}D_{\uparrow \downarrow
}(x_{mn}).  \label{eq2}
\end{equation}
The same formulas may be also obtained within the Landauer-B\"uttiker
scattering formalism \cite{Landauer,Buttik}. In the above expressions 
$\sigma ^F(\sigma ^{AF})$ is the conductance at ferromagnetic
(antiferromagnetic) alignment of domains, $\sigma _{\alpha \alpha }$ is the
conductance for the $\alpha $-th spin-channel, and $x_{mn}=\cos \theta $ is
the cosine of the quasiparticle incidence angle $\theta $, measured from the
longitudinal symmetry axis of the neck, indices $m$ and $n$ refer to quantum
numbers of transverse motion in the neck. $D_{\alpha \beta }(x)$ is the
quantum-mechanical transmission coefficient for the connecting channel (see
below). The magnetoresistance is defined as follows: 
\begin{equation}
MR=\frac{R^{AF}-R^F}{R^F}=\frac{\sigma ^F-\sigma ^{AF}}{\sigma ^{AF}}.
\label{eq3}
\end{equation}

Quantization of transverse motion in the channel obliges the parallel to the
interface projection of the incident quasiparticle momentum to satisfy the
requirement: 
\begin{equation}
p_{\parallel }=p_{F\alpha }\sin \theta =p_{mn}\equiv \hbar \lambda _{mn},
\label{eq4}
\end{equation}
where $p_{F\alpha }$ is the Fermi momentum for the $\alpha $-th
spin-channel, $\lambda _{mn}$ is the quantized wave number (see definition
below). This is the first basic selection rule, which comes from
quantization. Tilde in (\ref{eq1}) and (\ref{eq2}) means that the summations
should be done over the open conduction channels satisfying the condition: 
\begin{equation}
x_{mn}\equiv \cos \theta =\sqrt{1-(\hbar \lambda _{mn}/p_{F\alpha })^2}\leq
1.  \label{eq5}
\end{equation}
When the magnetizations alignment is ferromagnetic, the Fermi momenta on
both sides of the contact are equal in each, spin-up and spin-down, channel,
respectively. The energy and momentum conservations are satisfied, and the
transmission coefficients are equal to unity. At the antiferromagnetic
alignment the conservation of the parallel to the interface momentum 
($p_{\Vert }\equiv p_{F1\alpha }\sin \theta _1=p_{F2\alpha }\sin \theta _2$ ,
where the subscript 1 or 2 labels left- or right-hand side of the contact,
respectively) introduces the additional selection rule into Eq. (\ref{eq4}): 
\begin{equation}
p_{F\alpha }=\min (p_{Fj\uparrow },p_{Fj\downarrow }).  \label{eq6}
\end{equation}
This selection rule is strictly valid, if the electron spin conserves upon
transmission through the DW. We believe, that conservation is realized in
the atomic-size point contacts, when the length of the connecting channel is
comparable with the Fermi wave-length of the current carriers. It was argued
that the above scenario is realized, if the DW width $\delta <\delta _s$,
where $\delta _s=\min (v_F/\omega _Z,v_FT_1)$, $T_1$ is the longitudinal
relaxation rate time of the carriers magnetization \cite{Tagirov1}, and 
$\omega _z$ is the Larmore precession frequency \cite{Gregg}. Imamura 
\textit{et al.} \cite{Imamura} justified the above hypothesis for a quantum 
DW by numerical calculations for the linear chain of spins.

We perform concrete calculations for the neck of rectangular cross-section,
which models a contact with asymmetric cross-section. The solution to the
Schr\"odinger equation for the electron moving in the neck is sought in the
form 
\begin{equation}
\Psi (x,y,z)=\Phi (z)\sin \frac{\pi nx}a\sin \frac{\pi my}b.  \label{eq8}
\end{equation}
The function $\Phi (z)$ describes motion along the channel, it obeys the
equation 
\begin{equation}
\hbar ^2\frac{\partial ^2\Phi }{\partial z^2}+\left( p_{F0}^2-\lambda
_{mn}^2\hbar ^2+2MU(z)\right) \Phi =0,  \label{eq9}
\end{equation}
where $U(z)=zE_{ex}/l$ is the potential landscape created in the neck by the
constrained domain wall, $E_{ex}$ is the conduction band exchange energy
splitting, $M$ is the conduction electron mass, $a$ and $b$ are the width
and height of the neck and $p_{F0}$ is the Fermi momentum in absence of
conduction band splitting. In Eqs. (\ref{eq8}) and (\ref{eq9}) $m$ and $n$
are positive integer quantum numbers. The discrete function $\lambda _{mn}$
is given by 
\begin{equation}
\lambda _{mn}=\pi \sqrt{\left( \frac na\right) ^2+\left( \frac mb\right) ^2.}
\label{eq10}
\end{equation}
The choice of potential energy $U(z)$ in the form of linear function of $z$
is based on the calculations by Bruno (Ref. \cite{Bruno}, Fig. 2). Eq. (\ref
{eq9}) has an exact solution in terms of Airy functions, the explicit
expression for the transmission coefficient $D_{\alpha \beta }$ is given in
Ref. \cite{Tagirov2}.

A numerical routine consists of the summation over the consecutive values of
the roots $\lambda _{mn}$ satisfying the constraints, Eqs. (\ref{eq4}) and 
(\ref{eq5}). At the antiferromagnetic alignment the minority Fermi momentum
of the either spin projection should be used instead of $p_{F\alpha }$ in
Eqs. (\ref{eq4}) and (\ref{eq5}) to calculate the conductance $\sigma ^{AF}$, 
Eq. (\ref{eq2}). The results are displayed on Figures 1 and 2, important
for the discussion conduction band spin-polarization parameter $\gamma $ is
defined as: $\gamma =p_{F\downarrow }/p_{F\uparrow }\leq 1$. Calculations
revealed that the results depend on the absolute value of $p_{F\uparrow }$ ,
we have chosen $\hbar ^{-1}p_{F\uparrow }=1$\AA $^{-1}$ for the presentation.

\section{Results of calculations}

Fig.1 displays the results of calculations for the neck of the square
cross-section ($b=a$) and $\gamma =0.68$. Panel (a) shows the dependence 
of F- and AF-conductances on the channel radius. $l$ and $\lambda
=lp_{F\uparrow }\hbar ^{-1}$ are the length in \AA\ and dimensionless length
of the connecting channel, respectively. The chosen value, $\lambda =10.0$,
corresponds to the connecting channel length 10\AA\ (1 nm). Panel (b) shows
the dependence of magnetoresistance on the channel size $a$. The panels (c)
and (d) display the magnetoresistance against F-conductance for the sloping
(c) and the step-like (d) potential landscapes in the channel, the latter
one is the limiting case of very sharp DW. Physically, Fig.1 demonstrates
the case, when the AF-alignment conduction opens in the interior part of the
first F-conductance plateau. It allows us to make the following conclusions:
1) the F-alignment conductance is spin-dependent, the conductions of
spin-channels open asynchronously (panel (a)), thus resulting in $e^2/h$
quantization of conductance \cite{Imamura,Zvezdin}; 2) if some number of
conduction channels are open for the F-alignment ($\sigma ^F$ is finite),
but there is no conduction for the AF-alignment ($\sigma ^{AF}=0$), then,
according to definition Eq. (\ref{eq3}), MR diverges. Magnetoresistance is
infinite in the idealized model with no reversal of the carriers spin upon
transmitting the neck. In a more realistic treatment the magnitude of MR of
this quantum spin-valve is restricted by the spin-flip process, which gives
rise to a finite AF-conductance at any number of open F-conductance
channels. It is the quantum spin-valve regime; 3) the magnitude of MR beyond
the quantum spin-valve regime is well above 200\% for very moderate
polarization of the conduction band ($\gamma =0.68$, see discussion below);
4) the magnetoresistance has very sharp and high peak, when the first
channel for AF conductance step appears (panel (b) in correlation with
panel(a)); 5) sudden jumps in magnetoresistance, followed by practically
flat plateaus, appear at the moments when new F-alignment spin-up
conductance channel opens. They persist until the spin-down projection opens
new channel (panel (b) in correlation with panel (a)); 6) panels (c) and (d)
show that the magnetoresistance drawn as a function of quantized F-alignment
conductance is a multivalued function of F-conductance, $\sigma ^F$ \cite
{Tagirov2}. The issue 4) leads to weakly disperse, or even non-disperse
behavior of magnetoresistance at certain numbers of open F-alignment
channels: $N^F=4,5,7,11,13,17...$ (see panels (c) and (d)). Non-disperse
behavior of MR comes if the AF-conductance is practically independent on the
contact size when a new F-conductance channel opens (see panel (a)). The
issue 6) means, that if the temperature and disorder effects can be
neglected, several values of magnetoresistance correspond to the same number
of open conductance channels for the F-alignment of magnetizations (abscissa
in the panels (c) and (d)). The overall width of distributions of MR points,
which belong to the same value of the quantized F-conductance, may be
comparable with maximal value of MR, \textit{i.e.} magnetoresistance
acquires giant fluctuations because of conductance quantization.

Next, we change the aspect ratio $\varepsilon $ of the sides of the
rectangular cross-section, $\varepsilon =b/a.$ Fig. 2 is drawn with 
$\varepsilon =1.5$. Main changes compared to Fig. 1 can be summarized as
follows: 1) the AF-conductance opens now at three open channels of
F-conductance ($\sigma ^F=3e^2/h$, panels (c) and (d)); 2) the range of the
neck sizes with zero AF-conductance (quantum spin-valve) becomes broader; 3)
the overall magnitudes of MR increase (panels (b)-(d)); 4) magnetoresistance
points appear at almost every number of open F-conductance channels (compare
panels (c) and (d) of Figs. 1 and 2). We emphasize the issue which has
important implication to point contact GMR experiments: number of open
F-conductance channels, at which the AF conductance opens ($\sigma ^F=
2e^2/h$ in Fig.1 and $\sigma ^F=3e^2/h$ in Fig.2) depends not only on the
polarization of the conduction band \cite{Tagirov2}, but also on the
asymmetry of the point contact cross-section.

\section{Discussion of the results}

There are techniques, which provide the information about the
spin-polarization of the ferromagnet's conduction band at the Fermi energy.
These are the ferromagnet-insulator-superconductor tunneling spectroscopy
(see Ref. \cite{TedMes} and references therein) and the Andreev-reflection
spectroscopy \cite{Soulen98,Soulen99,Buhrman,Nadgorny,Strijk1,Strijk2}. The
tunneling spectroscopy suggests the following estimates for the mean values
of conduction band polarization parameter $\gamma $: 0.6 for permalloy 
(Ni$_{80}$Fe$_{20}$); 0.63 for pure Ni; 0.48 for Co and 0.43 for Fe. From the
Andreev-reflection spectroscopy we obtain the ranges for the values of 
$\gamma $: $\sim 0.68$ for permalloy; $\sim 0.62-0.72$ for Ni; $\sim 0.6-0.68$
for Co; $\sim 0.62-0.64$ for Fe. Observing the tunneling and
Andreev-reflection data on $\gamma $ and our figures we may confirm our
conclusion made from calculations for the cylindrical neck in Ref. \cite
{Tagirov2}: using realistic values of $\gamma $ we may reproduce maximal
values as well as giant fluctuations of MR data obtained by Garc\'ia 
\textit{\ et al.} \cite{Garcia1,Garcia2,Garcia3}. However, the agreement 
between the theory and the experiment on Fig. 3 in \cite{Tagirov2} could be 
even better, if some MR points would appear at neighboring number of open 
F-conductance channels. Comparison of Figs. 1 and 2 of the present work with 
Fig. 3 from Ref. \cite{Tagirov2} shows, that varying the aspect ratio in the range 
$\sim 1.0-2.0$ one may get a desired re-assignment of some MR points to number 
of conduction channels, and to improve agreement between the theory and the
experiment.

Independent on the actual shape of the neck, when its length is comparable
or longer than the cross-section size, the dipole-dipole anisotropy energy
may cause fluctuations between Bloch, N\'eel or more complicated types of
domain walls. Coey \textit{et al.} concluded \cite{Coey1,Coey2} that giant
MR of a nanocontact may be reduced somewhat by these fluctuations, but not
eliminated. In recent calculations Zhuravlev \textit{et al.} \cite{Zhurav1}
also predicted giant values and fluctuations of MR in segmented nanowires,
when conductance of the wire is quantizes.

When the cross-section of the point contact is very small, so that
F-conduction is open for only one spin-channel, the magnitude of GMR is
limited from above by the spin-reversal rate of conduction electrons upon
passing through the neck \cite{Tagirov2}. Our calculations show (see panels
(a) of Figs. 1 and 2), that higher the polarization of the conduction band
and larger the asymmetry of the cross-section, then wider the range of neck
sizes, at which the regime of quantum spin-valve can be realized. Magnetic
half-metal contacts with $\sim $100\% polarization of conduction band would
be almost always quantum spin-valves at nanometer range of size. In very
recent experiments \cite{Chopra,Garcia4} the ballistic magnetoresistance
(BMR) in the range 3000-4000\% has been observed in Ni point contact. These
really giant MR values can be easily reproduced in the ballistic regime of
quasiclassical conductance, Eq. (23) of Ref. \cite{Tagirov1}, for a moderate
polarization of the conduction band: with $\gamma =0.2$ ($P$(DOS$)=100\cdot
(1-\gamma )/(1+\gamma )=67\%$) we get $MR=3090\%$, and with $\gamma 
=0.18$ ($P=70\%$) we get $MR=4140\%$. However, Garc\'\i a \textit{et al.} 
\cite{Garcia4} reported also in the footnote, Ref. 9, that few times GMR up to
100000\% was observed in magnetic nanocontacts. Concerning this information,
we may guess that this huge magnetoresistance could be actually the result
of the quantum spin-valve realization. In contrast to the explanation
proposed in \cite{Garcia4}, the quantum spin-valve hypothesis does not need
in almost completely (100\%) polarized conduction band to predict 100000\%
effect. Theoretically, these numbers may appear even at experimentally
approved polarizations of Ni conduction band in the range 35-45\% \cite
{TedMes,Soulen98,Soulen99,Buhrman,Nadgorny,Strijk1}. It seems, that quantum
spin-valve concept brings us to the upper physical limit of
magnetoresistance for a non-superconducting spin-valve-type device. The true
infinite (but positive) magnetoresistance can be reached in the
proximity-effect superconducting spin-valve (PRESUS-valve) proposed in \cite
{Tagirov3,Buzdin}.

In conclusion, we have investigated theoretically the giant
magnetoresistance of a nanosize magnetic point contact in the regime of
conductance quantization. Concrete calculations have been made for the neck
of rectangular cross-section, and dependence of GMR on the asymmetry of
cross-section has been studied. Results of calculations show that taking
into consideration possible asymmetry of the point contact cross-section one
may improve agreement between the theory and the experiment. We argued, that
if conductance is open for only one spin-channel, the MR magnitude of this
quantum spin-valve is limited by the spin-reversal rate of conduction
electrons. For larger areas of the nanocontact the magnetoresistance becomes
a multivalued function of the conductance $\sigma ^F$ at ferromagnetic
alignment of contacting magnetic domains. This multivalued behavior of MR
(which may be treated as giant reproducible fluctuations of MR) is the
intrinsic property of quantum magnetic nanocontacts. This property survives
for every shape of the nanocontact and disorder, provided that: 1)
conductance at the ferromagnetic alignment is quantized (steps are not
destroyed); 2) the domain wall in the constriction is effectively sharp.
When observed experimentally, such MR distributions should not be
interpreted as being due to poor reliability or reproducibility of
experimental data.

\section{Acknowledgments}

The authors have benefited from collaboration with Prof. K.B. Efetov. The
work is supported by URFI grant N$^0$ 01.01.061 and by BRHE grant REC-007.

%\newpage

\begin{center}
\textbf{Figure captions}
\end{center}

Fig. 1. The dependence of conductance (a), and MR (b) on the cross-sectional
size of the neck $a$, $\varepsilon =1.0$. Panels (c) and (d) show
dependencies of MR on the number of the open conductance channels at the
F-alignment of the magnetizations. The maximal MR=563\% for the step-like
potential at $\sigma ^F=2e^2/h$ is not shown.

Fig. 2. The same as in Fig.1, but for $\varepsilon =1.5$. The maximal
MR=758\% for the step-like potential and MR=322\% for the sloping potential
at $\sigma ^F=3e^2/h$ are not shown.


\begin{thebibliography}{99}
\bibitem{Wees}  B.J. van Wees, H. van Houten, C.W.J. Beenakker, J.G.
Williamson, L.P. Kouwenhoven, D. van der Marel, C.T. Foxton, Phys. Rev.
Lett. \textbf{60} (1988) 848.

\bibitem{Wharam}  D.A. Wharam, T.J. Thornton, R. Newbury, M. Pepper, H.
Ahmed, J.E.F. Frost, D.G. Hasko, D.C. Peacock, D.A. Ritchie, G.A.C. Jones,
J. Phys. C \textbf{21} (1988) L209.

\bibitem{Kramer}  J.L. Costa-Kr\"amer, Phys. Rev. B \textbf{55} (1997) 4875.

\bibitem{Oshima}  H. Oshima, K. Miyano, Appl. Phys. Lett. \textbf{73 }(1998)
2203.

\bibitem{Ott}  F. Ott, S. Barberan, J.G. Lunney, J.M.D. Coey, P. Berthet,
A.M. de Leon-Guevara, A. Revcolevschi, Phys. Rev. B \textbf{58} (1998) 4656.

\bibitem{Ono}  T. Ono, Y. Ooka, H. Miyajima, Appl. Phys. Lett. \textbf{75}
(1999) 1622 .

\bibitem{Imamura}  H. Imamura, N. Kobayashi, S. Takahashi, S. Maekawa, 
Phys. Rev. Lett. \textbf{84} (2000) 1003.

\bibitem{Zvezdin}  A.K. Zvezdin, A.F. Popkov, JETP Lett. \textbf{71} (2000)
209.

\bibitem{Garcia1}  N. Garc\'ia, M. Mu\~noz, Y.-W. Zhao, Phys. Rev. Lett. 
\textbf{82} (1999) 2923.

\bibitem{Garcia2}  G. Tatara, Y.-W. Zhao, M. Mu\~noz, N. Garc\'ia, Phys.
Rev. Lett. \textbf{83} (1999) 2030.

\bibitem{Garcia3}  N. Garc\'ia, M. Mu\~noz, Y.-W. Zhao, Appl. Phys. Lett. 
\textbf{76} (2000) 2586.

\bibitem{Tagirov2}  L.R. Tagirov, B.P. Vodopyanov, K.B. Efetov, Phys. Rev. B 
\textbf{65} (2002) 214419.

\bibitem{Referee}  We acknowledge the referee of the work \cite{Tagirov2}
who stimulated the analysis of the one-spin-projection conductance regime.

\bibitem{Falicov}  G.G. Cabrera, L.M. Falicov, Phys. Stat. Solidi B 
\textbf{61} (1974) 539.

\bibitem{Hubert}  A. Hubert, R. Sch\"afer, \emph{Magnetic Domains}
(Springer, Berlin, 1988).

\bibitem{Bruno}  P. Bruno, Phys. Rev. Lett. \textbf{83} (1999) 2425.

\bibitem{Tagirov1}  L.R. Tagirov, B.P. Vodopyanov, K.B. Efetov, Phys. Rev. B 
\textbf{63} (2001) 104428.

\bibitem{Levy}  P.M. Levy, Sh. Zhang, Phys. Rev. Lett. \textbf{79} (1997)
5110.

\bibitem{Landauer}  R. Landauer, IBM J. Res. Dev. \textbf{32} (1988) 306.

\bibitem{Buttik}  M. B\"uttiker, IBM J. Res. Dev. \textbf{32} (1988) 317.

\bibitem{Gregg}  J.F. Gregg, W. Allen, K. Ounadjela, M. Viret, M. Hehn, S.M.
Thompson, J.M.D. Coey, Phys. Rev. Lett. \textbf{77} (1996) 1580.

\bibitem{TedMes}  P.M. Tedrow and R. Meservey, Phys. Rep. \textbf{238}
(1994) 173.

\bibitem{Soulen98}  R.J. Soulen Jr., J.M. Byers, M.S. Osofsky, B. Nadgorny,
T. Ambrose, S.F. Cheng, P.R. Broussard, C.T. Tanaka, J. Nowack, J.S.
Moodera, A. Barry, J.M.D. Coey, Science \textbf{282} (1998) 85.

\bibitem{Soulen99}  R.J. Soulen Jr., M.S. Osofsky, B. Nadgorny, T. Ambrose,
P. Broussard, S.F. Cheng, J. Byers, C.T. Tanaka, J. Nowack, J.S. Moodera, G.
Laprade, A. Barry, J.M.D. Coey, J. Appl. Phys. \textbf{85} (1999) 4589.

\bibitem{Buhrman}  S.K. Upadhyay, A. Palanisami, R.N. Louie, R.A. Buhrman,
Phys. Rev. Lett. \textbf{81} (1998) 3247.

\bibitem{Nadgorny}  B. Nadgorny, R.J. Soulen, M.S. Osofsky, I.I. Mazin, G.
Laprade, R.J.M. van de Veerdonk, A.A. Smits, S.F. Chang, E.F. Skelton, S.B.
Qadri, Phys. Rev. B \textbf{61} (2000) 3788.

\bibitem{Strijk1}  G.J. Strijkers, Y. Ji, F.Y. Yang, C.L. Chien, J.M. Byers,
Phys. Rev. B \textbf{63} (2001) 104510.

\bibitem{Strijk2}  Y. Ji, G.J. Strijkers, F.Y. Yang, C.L. Chien, J.M. Byers,
A. Anguelouch, G. Xiao, A. Gupta, Phys. Rev. Lett. \textbf{86} (2001) 5585.

\bibitem{Coey1}  J.M.D. Coey, L. Berger, Y. Labaye, Phys.Rev. B \textbf{64}
(2001) 020407.

\bibitem{Coey2}  Y. Labaye, L. Berger, J.M.D. Coey, J. Appl. Phys. 
\textbf{91} (2002) 5341.

\bibitem{Zhurav1}  M.Ye. Zhuravlev, H.O. Lutz, A.V. Vedyayev, Phys. Rev. B 
\textbf{63} (2001) 174409.

\bibitem{Chopra}  H.D Chopra, S.Z. Hua, Phys. Rev. B \textbf{66} (2002)
020403(R).

\bibitem{Garcia4}  H. Wang, H. Cheng, N. Garc\'\i a, cond-mat/0207516 (22
July 2002).

\bibitem{Tagirov3}  L.R. Tagirov, Phys. Rev. Lett. \textbf{83} (1999) 2058.

\bibitem{Buzdin}  A.I. Buzdin, A.V. Vedyayev, N.V. Ryzhanova, Europhys.
Lett. \textbf{48} (1999) 686.
\end{thebibliography}
\end{document}